\documentclass[,final]{aipproc}

\layoutstyle{8x11single}

\usepackage{fancyhdr}
\usepackage{amssymb,amsmath}
\usepackage{tikz}
\usepackage{graphicx}
\usepackage{slashed}
\usepackage{soul}

\newcommand{\eq}[1]{Eq.~(\ref{#1})}

\newcommand{\eV}{\mathinner{\mathrm{eV}}}
\newcommand{\keV}{\mathinner{\mathrm{keV}}}

\newcommand{\GeV}{\mathinner{\mathrm{GeV}}}
\newcommand{\TeV}{\mathinner{\mathrm{TeV}}}

\def\l{\left}
\def\r{\right}
\def\pl{{\rm P}}

\def\bea{\begin{eqnarray}}
\def\eea{\end{eqnarray}}
\def\beq{\begin{equation}}
\def\eeq{\end{equation}}

\begin{document}

\title{An alternative to the standard model}


\author{Seungwon Baek, Pyungwon Ko and Wan-Il Park}{
  address={School of Physics, KIAS, Seoul 130-722, Korea}
}

%

\begin{abstract}
We present an extension of the standard model to dark sector with an unbroken local dark $U(1)_X$ symmetry.
Including various singlet portal interactions provided by the standard model Higgs, right-handed neutrinos and kinetic mixing, we show that the model can address most of phenomenological issues (inflation, neutrino mass and mixing, baryon number asymmetry, dark matter, direct/indirect dark matter searches, some scale scale puzzles of the standard collisionless cold dark matter, vacuum stability of the standard model Higgs potential, dark radiation) and be regarded as an alternative to the standard model.
The Higgs signal strength is equal to one as in the standard model for unbroken 
$U(1)_X$ case with a scalar dark matter,  but it could be less than one independent of 
decay channels if the dark matter is a dark sector  fermion or if $U(1)_X$ is spontaneously 
broken, because of a mixing with  a new neutral scalar boson in the models.  
\end{abstract}

\maketitle


\section{Introduction}

Even though its unsurpassed success in describing subatomic world, it is well known that the standard model (SM) has to be extended in order to accommodate ($i$) neutrino masses and mixings, ($ii$) baryon number asymmetry, ($iii$) nonbaryonic cold dark matter 
of the universe, and ($iv$) inflation and density perturbation. 
For the first two items, the most economic and aesthetically attractive idea is to introduce singlet right-handed neutrinos and the seesaw mechanism, and leptogenesis~\cite{leptogenesis} for baryon number asymmetry.  
For the 3rd item, there are many models for cold dark matter (CDM), from axion to lightest superparticles and hidden sector DMs, to 
name a few.  
For the 4th item the simplest inflation model without new inflaton fields would be $R^2$ inflation by Starobinsky~\cite{R2} and Higgs 
inflation~\cite{Bezrukov:2007ep}. 

One of the puzzles in the CDM physics is how CDM can be absolutely stable or very long lived. 
The required longevity ($10^{26-30} {\rm sec}$ \cite{Ackermann:2012qk}) of the dark matter (DM) can be guaranteed by a symmetry.
In nonsupersymmetric dark matter models, one often assumes ad hoc $Z_2$ symmetry in order to stabilize DM, without deeper understanding of its origin or asking if it is global or local discrete symmetry.  
If we assume that global symmetry is not protected by quantum gravity effects, this $Z_2$ symmetry would be broken by $1/M_{\rm Planck}$ suppressed nonrenormalizable operators with an order unity dimensionless numerical coefficient~\cite{Baek:2013qwa}. 
Then the electroweak scale DM can not live long enough to be dark matter candidate of the universe. 
The simplest way to guarantee the stability of electroweak (EW) scale DM is to assume the DM carries its own gauge charge which is absolutely conserved. 
Thus we are led to local dark symmetry and dark gauge force. 
This would be a very natural route for the DM model building, since the unsurpassed successful SM is also based on local gauge symmetry and its spontaneous breaking.
A local symmetry other than the SM gauge group often appears in theories beyond the SM, and would guarantee the absolute stability of dark matter if it is unbroken or broken to a $Z_2$ discrete symmetry as the remnant of the symmetry breaking.

If dark gauge coupling is strong and dark gauge interaction is confining like ordinary QCD, the DM would be the lightest composite hadrons in the hidden sector.  
In this case it is possible to generate all the mass scales of the SM particles as well as the DM mass from dimensional transmutation in the hidden sector strong interaction~\cite{Hur:2011sv}. 
If dark gauge coupling is weak, we can employ the standard perturbation method to analyze the problems, which we adopt in the model described in this talk.

Another guiding principle of extending SM is renormalizability of the model. 
The present authors found that one would get erroneous results if the effective Lagrangian approach is used for singlet fermion or vector DM with Higgs portal~\cite{Baek:2011aa,Baek:2012se}.  

Dark sector can communicate with the SM sector via Higgs portal interactions ($H^\dagger H$) which are quite often used in the dark matter physics. 
Another possible renormalizable portal interaction can be provided by heavy RH neutrinos
which are singlet under  the SM gauge group. 
These singlet portal interactions are natural extensions of the SM in the framework of  renormalizable quantum field theory,  and allow rich phenomenology in both dark matter and Higgs sector as we will show in the subsequent sections.

In this talk, we present a simple renormalizable model where the dark matter lives in a dark (hidden) sector with its own dark gauge charge along with dark gauge force. 
We mainly discuss the unbroken $U(1)_X$ dark gauge symmetry, and briefly mention what happens if $U(1)_X$ is spontanesouly broken. 
This talk is based on Ref.~\cite{Baek:2013qwa}, to which we invite the readers for more detailed discussions on the subjects described in this talk.



\section{The Model}

We assume that dark matter lives in a hidden sector, and it is stable 
due to unbroken local $U(1)_X$ dark gauge symmetry.  All the SM fields are taken to be $U(1)_X$ 
singlets. Assuming that the RH neutrinos are portals to the hidden sector, we need both a scalar ($X$) 
and a Dirac fermion ($\psi$) with the same nonzero dark charge (see Table~1). 
Then the composite operator $\psi X^\dagger$ becomes a gauge singlet and thus can 
couple to the  RH neutrinos $N_{Ri}$'s.  

With these assumptions, we can write the most general renormalizable Lagrangian as follows:   
\beq \label{Lagrangian}
\mathcal{L} = \mathcal{L}_{\rm SM} + \mathcal{L}_X + {\mathcal{L}_\psi} 
+ \mathcal{L}_{\rm kin-mix} +  \mathcal{L}_{\rm H-portal} + \mathcal{L}_{\rm RHN-portal}  
\eeq
where $\mathcal{L}_{\rm SM}$ is the standard model Lagrangian and  
\bea
\label{LX}
\mathcal{L}_X &=& {\l| \l( \partial_\mu + i g_X q_X \hat{B}'_\mu \r) X \r|^2} - \frac{1}{4} \hat{B}'_{\mu \nu} \hat{B}'^{ \mu \nu} - m_X^2 X^\dag X - \frac{1}{4} \lambda_X \l( X^\dag X \r)^2 ,
\\
\mathcal{L}_\psi &=& i \bar{\psi} \gamma^\mu \l( \partial_\mu + i g_X q_X \hat{B}'_\mu \r) \psi - m_\psi \bar{\psi} \psi ,
\\
\label{kin-mix}
\mathcal{L}_{\rm kin-mix} &=& - \frac{1}{2} \sin \epsilon \hat{B}'_{\mu \nu} \hat{B}^{\mu \nu} ,
\\
\label{Hportal}
\mathcal{L}_{\rm H-portal} &=& - \frac{1}{2} \lambda_{HX} X^\dag X H^\dag H ,
\\
\label{RHNportal}
- \mathcal{L}_{\rm RHN-portal} &=& \frac{1}{2} M_i \overline{N_{Ri}^C} N_{Ri} + \left[ Y_\nu^{ij} \overline{N_{Ri}} \ell_{Lj} H^\dag + \lambda^i \overline{N_{Ri}} \psi X^\dag + \textrm{H.c.} \right]. 
\eea
$g_X$, $q_X$, $\hat{B}'_\mu$ and $\hat{B}'_{\mu \nu}$ are the gauge coupling, $U(1)_X$ charge, the gauge field and 
the field strength tensor of the dark $U(1)_X$, respectively. 
$\hat{B}_{\mu \nu}$ is the gauge field strength of the SM $U(1)_Y$.
We assume 
\beq
m_X^2 > 0, \quad \lambda_X > 0, \quad \lambda_{HX} > 0
\eeq
so that the local $U(1)_X$ remains unbroken and the scalar potential is bounded from below 
at tree level.

Either $X$ or $\psi$ is absolutely stable due to the unbroken local $U(1)_X$ gauge 
symmetry, and will be responsible for  the present relic density of nonbaryonic CDM.  
In our model, there is a massless dark photon which couples to the  SM $U(1)_Y$ gauge field 
by kinetic mixing.  After diagonalization of the kinetic terms
the SM $U(1)_Y$ gauge coupling is redefined as 
$g_Y = \hat{g}_Y / \cos \epsilon$, and hidden photon does not couple to the SM fields.
However, dark sector fields now couple to the SM photon and $Z$-boson.
In the small mixing limit, the couplings are approximated to 
\beq \label{DM-SM-int}
\mathcal{L}_{\rm DS-SM} = {{\bar{\psi} i \gamma^\mu \l[ \partial_\mu - i g_X q_X t_\epsilon \l( c_W A_\mu - s_W Z_\mu \r) \r] \psi}} + \l| \l[ \partial _\mu - i g_X q_X t_\epsilon \l( c_W A_\mu - s_W Z_\mu \r) \r] X \r|^2
\eeq
where $t_\epsilon = \tan \epsilon$, $c_W = \cos \theta_W$ and $s_W = \sin \theta_W$ with $\theta_W$ being the Weinberg angle.
Hence, dark sector fields charged under $U(1)_X$ can be regarded as mini-charged particles under electromagnetism after the kinetic mixing term is removed by a field redefinition,  Eq.~(2.8). 

Meanwhile, we can assign lepton number ($q_L$) and $U(1)_X$ charge ($q_X$) to RH neutrinos and dark fields 
as $(q_L, q_X)_i = (1,0)_N, \ (1,1)_\psi, \ (0,1)_X$.
Then, the global lepton number is explicitly broken by Majorana mass terms for the RH neutrinos.
If $Y_\nu$ and $\lambda_i$ carry $CP$-violating phases, the decay of RH neutrinos can develop 
lepton number asymmetry in both of visible and dark sectors.  
Since $U(1)_X$ is unbroken, the net charge of the dark sector should be zero though each of dark sector fields can have non-zero charge asymmetry when they coexist.

There are various physics issues involved in our model as listed below: 
\begin{itemize}
\item Small and large scale structure
\item Vacuum stability of Higgs potential
\item CDM relic density and direct/indirect DM searches
\item Dark radiation
\item Leptogenesis
\item Higgs inflation in case of a large non-minimal gravitational couplings
\end{itemize}
Although the model is highly constrained, it turns out that it can explain various issues related to those physics in its highly constrained narrow parameter space without conflicting with any phenomenological, astrophysical and cosmological observations. 

\section{Constraints}
In this section, we will take a look each of constraint or physics in our model one by one.

\subsection{Structure formation}
\label{structure-form}
The presence of the dark matter self-interaction caused by nonzero charge of $U(1)_X$ could affect significantly the kinematics, shape and density  profile of dark matter halo, so it is constrained by, for example, the galactic dynamics \cite{Ackerman:2008gi}, ellipticity of dark matter halos \cite{MiraldaEscude:2000qt} and Bullet Cluster \cite{Randall:2007ph}. 
In the case of a velocity-dependent self-interaction, the transfer cross section of the dark matter self-interaction is upper-bounded as \cite{Vogelsberger:2012ku}
\beq \label{sigmaT-obs}
\l. \frac{\sigma_T^{\rm obs}}{m_{\rm dm}} \r|_{v=10 {\rm km/s}} \lesssim 35 \ {\rm cm^2/g}.
\eeq 
Interestingly, it was shown that, if $\sigma_T^{\rm obs}$ is close to the bound, it can solve the core/cusp problem \cite{Oh:2010ea} and ``too big to fail'' problem \cite{BoylanKolchin:2011dk} of the standard collisionless CDM scenario \cite{Vogelsberger:2012ku}.  

In our model, for both $\psi$ and $X$ the self-interaction cross section with a massless dark photon is given by \cite{Feng:2009mn} 
\beq
\sigma_T \simeq \frac{16 \pi \alpha_X^2}{m_{X(\psi)}^2 v^4} 
\ln \l[ \frac{m_{X(\psi)}^2 v^3}{(4 \pi \rho_X\alpha_X^3)^{1/2}} \r]
\eeq
where $v$ and $\rho_X$ are the velocity and density of the dark matter at the region of interest.
We take $v = 10 {\rm km/sec}$ and $\rho_X = 3 \GeV /{\rm cm}^3$.
Then, compared to \eq{sigmaT-obs}, dark interaction is constrained as
\beq \label{DMstructure-const}
\alpha_X \lesssim 5 \times 10^{-5} \l( \frac{m_{X(\psi)}}{300 \GeV} \r)^{3/2}
\eeq
where we approximated the log factor to $41$.
Either $X$ or $\psi$, which is lighter than the other, poses a stronger constraint on $\alpha_X$.

\subsection{CDM relic density}
$\psi$ couples only to dark photon at low energy, and the thermally-averaged  
annihilation cross section of 
$\psi$ is found to be $\langle \sigma v \rangle_{\rm ann}^\psi \approx \pi \alpha_X^2 / (2 m_\psi^2)$.
Hence, for $\alpha_X$ satisfying \eq{DMstructure-const}, $\psi$ would be over-abundant at present if stable.
In order to avoid the over-closing by $\psi$, we assume
\beq \label{CDM-const}
m_\psi > m_X
\eeq
so that $\psi$ can decay through the virtual RH neutrinos.
The decay rate of $\psi$ is given by 
\beq
\Gamma_\psi \simeq \frac{\lambda_1^2}{16 \pi} \frac{\tilde{m}_\nu}{M_1} m_\psi \l( 1 - \frac{m_X^2}{m_\psi^2} \r)^2 \l[ 1 + \frac{1}{48 \pi^2} \l( \frac{m_\psi^2}{v_H^2} \r) \r]
\eeq
where $\tilde{m}_\nu \equiv Y_\nu^2 v_H^2 / M_1$ and $v_H=174 \GeV$ are respectively a contribution to the neutrino mass matrix and the vev of Higgs field. 

%
If $\psi$ decays before the thermal component of $X$ freezes out, the $X$'s coming 
from the decay of $\psi$ thermalize, which makes the number density $n_X$ return to 
that of thermal equilibrium.  The present relic density in this case is determined by the 
thermal relic.
On the other hand, if $\psi$ decays after the thermal freeze-out of $X$, the annihilation cross section should be larger than the usual value for thermal relics 
so that the non-thermal freeze-out to provide a right amount of relic density (``symmetric non-thermal'' case).
In this case, the required background temperature when $\psi$ decays is determined by the annihilation cross section of $X$.

The pair annihilation of $X$-$X^\dag$ is controlled by the Higgs portal interaction $\lambda_{HX}$ which leads to $s$-wave annihilations. 
It freezes out at a temperature $T_f \sim m_X / 20$.
However dark matter can still be in kinetic equilibrium with thermal background at a lower temperature due to scatterings to SM particles.   
The scattering is mediated by photon and Higgs thanks to the kinetic mixing and Higgs portal interaction.
The transfer cross section of photon-mediation is such that 
$\sigma_T \propto \epsilon^2 / T^2$.
Although the associated scattering could be quite efficient at a low temperature, we found that, for $\epsilon \sim \mathcal{O}(10^{-9})$ which will be of our interest as described in section~3.3, the momentum transfer rate via photon is too small to keep kinetic equilibrium after freeze-out. 
In case of Higgs mediation, the kinetic equilibrium can be maintained to a temperature of the charm quark mass scale \cite{Hofmann:2001bi}.
If $\psi$ decays to $X$ abundantly, $X$ and $X^\dag$ would be able to re-annihilate even after freeze-out of the thermal annihilation until their number densities is reduced enough to stop the re-annihilation.
The abundance of $X$ and $X^\dag$ at the freeze-out of the re-annihilation should be responsible for the present relic density of dark matter.   
Equating the annihilation rate to the expansion rate when $\psi$ decay and setting the abundance of dark matter ($n_X/s$) is equal to the observed value, we find the decay temperature of $\psi$ to give a right amount of non-thermal relic density,
\beq
T_{\rm d} \equiv \l( \frac{\pi^2}{90} g_*(T_{\rm d}) \r)^{-1/4} \sqrt{\Gamma_\psi M_\pl} \simeq 9.2 \GeV \l( \frac{m_X}{300 \GeV} \r) \l( \frac{\langle \sigma v \rangle_{\rm ann}^{\rm th}}{\langle \sigma v \rangle_{\rm ann, d}^X} \r)
\eeq
where $\langle \sigma v \rangle_{\rm ann,d}^X$ is the annihilation cross section of $X$ when $\psi$ decays, and we used $g_*(T_{\rm d})=g_{*S}(T_{\rm d})=100$ and $M_\pl = 2.4 \times 10^{18} \GeV$ in the right-hand side of above equation.
This implies that 
\beq \label{lambda1-CDM-const}
\lambda_1^2 
\simeq 58 \l( \frac{0.1 \eV}{\tilde{m}_\nu} \r) \l( \frac{M_1}{10^9 \GeV} \r) \l( \frac{1 \TeV}{m_\psi} \r) \l( \frac{m_\psi^2}{m_\psi^2- m_X^2} \r)^{2} \l[ 1 + \frac{\l( m_\psi/v_H \r)^2}{48 \pi^2} \r]^{-1} 
\l( \frac{m_X}{300 \GeV} \r)^2 \l( \frac{\langle \sigma v \rangle_{\rm ann}^{\rm th}}{\langle \sigma v \rangle_{\rm ann,d}^X} \r)^2 .
\eeq
Note that, if $\langle \sigma v \rangle_{\rm ann,d}^X = \langle \sigma v \rangle_{\rm ann}^{\rm th}$, $T_{\rm d}$ equal to or larger than $T_{\rm f}$ and corresponding $\lambda_1$ are fine. 
Note also that for the $\GeV$-scale kinetic decoupling temperature ($T_{\rm kd}$), $T_{\rm d} > T_{\rm kd}$ unless $\langle \sigma v \rangle_{\rm ann,d}^X$ is larger than$\langle \sigma v \rangle_{\rm ann}^{\rm th}$ by at least two orders of magnitude.
However, as described in section~\ref{dd}, only $\langle \sigma v \rangle_{\rm ann}^X / \langle \sigma v \rangle_{\rm ann} ^{\rm th} \lesssim 5$ is allowed, so we can take $\langle \sigma v \rangle_{\rm ann,d}^X = \langle \sigma v \rangle_{\rm ann}^X$.


\subsection{Vacuum stability}
\label{vac-stab}
In the standard model, Higgs potential becomes unstable at an intermediate scale because of 
top loop contributions to the Higgs quartic couplings, though it depends on some of the standard 
model parameters, for example top pole mass and strong interaction \cite{Alekhin:2012py}.
Such instability can be cured if Higgs field couples to other scalar field(s) \cite{Lebedev:2012zw,EliasMiro:2012ay,Baek:2012uj}.  Depending on the existence of mixing between Higgs and additional scalar(s), 
tree-level and/or loop effects should be able to remove the instability.  
In our model, $X$ does not develop non-zero VEV, and the SM Higgs is not mixed with $X$. In this case, 
the loop-effect should be large enough to remove the vacuum instability of the SM Higgs potential.
Note that the Dirac neutrino mass terms also contribute to the RG-running of the Higgs quartic coupling.
However it is a negative contribution reflecting the fermionic nature of the right-handed neutrinos \cite{Rodejohann:2012px}.
Hence, in order not to make worse the vacuum instability up to Planck scale, we take $Y_{\nu}^{ij} \lesssim 0.1$,
and ignore its contribution to the RG equation of Higgs quartic coupling.
Solving 2-loop RGEs for SM couplings and 1-loop RGEs for non-SM couplings numerically, we found that the vacuum stability of Higgs potential and perturbativity of the couplings require 
\beq
0.2 \lesssim \lambda_{HX} \lesssim 0.6, \quad 0 < \lambda_X \lesssim 0.2.
\eeq

\subsection{Direct detection}
\label{dd}
In our model, dark matter couples to the SM particles via neutral SM gauge bosons (see \eq{DM-SM-int}) and Higgs portal, hence both type of interactions provide channels for dark matter direct searches.
In the case of gauge boson exchange, the spin-independent (SI) dark matter-nucleon scattering cross section via photon exchange provides a strong constraint on the kinetic mixing.
For a scattering to a target atom with atomic number $Z$, the differential cross section of the Rutherford scattering of our dark matter  with respect to the recoil energy $E_{\rm r}$ is given by
\beq \label{dsdE-th}
\l. \frac{d \sigma_A}{d E_{\rm r}} \r|_{\rm th}
= \frac{2 \pi \epsilon_e^2 \alpha_{\rm em}^2 Z^2}{m_A E_{\rm r}^2 v^2} \mathcal{F}_A^2(E_r).
\eeq
where $\epsilon_e = - (g_X/e) q_X c_W \tan \epsilon$ is the electric charge of our dark matter, $m_A$ is the mass of the atom, $v$ is the lab velocity, and $\mathcal{F}_A(E_{\rm r})$ is the form factor of the target atom.
Experimentally, for the SI dark matter-nucleus scattering, the differential cross section with respect to the nucleus recoil energy is parameterized as
\beq \label{dsdE-exp}
 \l. \frac{d \sigma_A}{d E_{\rm r}} \r|_{\rm exp} = \frac{2 m_A Z^2}{\mu_p^2 v^2} \l( \sigma_p^{\rm SI} \r)_{\rm exp} \mathcal{F}_A^2(E_{\rm r})
\eeq
where $\mu_p = m_X m_p / \l( m_X + m_p \r)$ is the reduced mass of dark matter-proton system, $\l( \sigma_p^{\rm SI} \r)_{\rm exp}$ is the dark matter-proton scattering cross section constrained by experiments.   Note that the velocity dependence of \eq{dsdE-th} is 
the same as that of \eq{dsdE-exp}, and $\mathcal{F}^2_A(E_{\rm r}) / E_{\rm r}^2$ is a monotonically decreasing function for the range of $E_{\rm r}$ relevant in various direct detection experiments \cite{Lewin:1995rx}.
Hence, the kinetic mixing is  bounded from above as 
\beq \label{t-epsilon-bnd}
t_\epsilon < \l[ \frac{1}{\pi q_X^2 c_W^2 \alpha_X \alpha_{\rm em}} \r]^{1/2} \l( \frac{m_A}{\mu_p} \r) E_{\rm r}^{\rm T} \l( \sigma_p^{\rm SI} \r)^{1/2}
\eeq
where $E_{\rm r}^{\rm T}$ is the threshold recoil energy of a target atom at a given experiment.

\begin{figure}[ht] 
\centering
\includegraphics[width=0.45\textwidth]{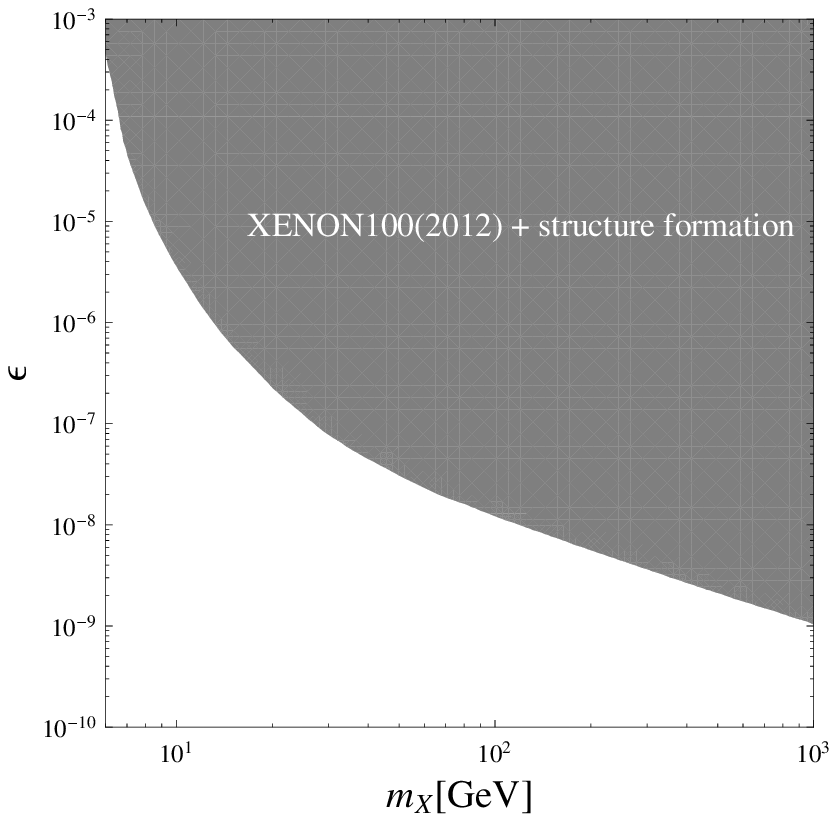}
\includegraphics[width=0.45\textwidth]{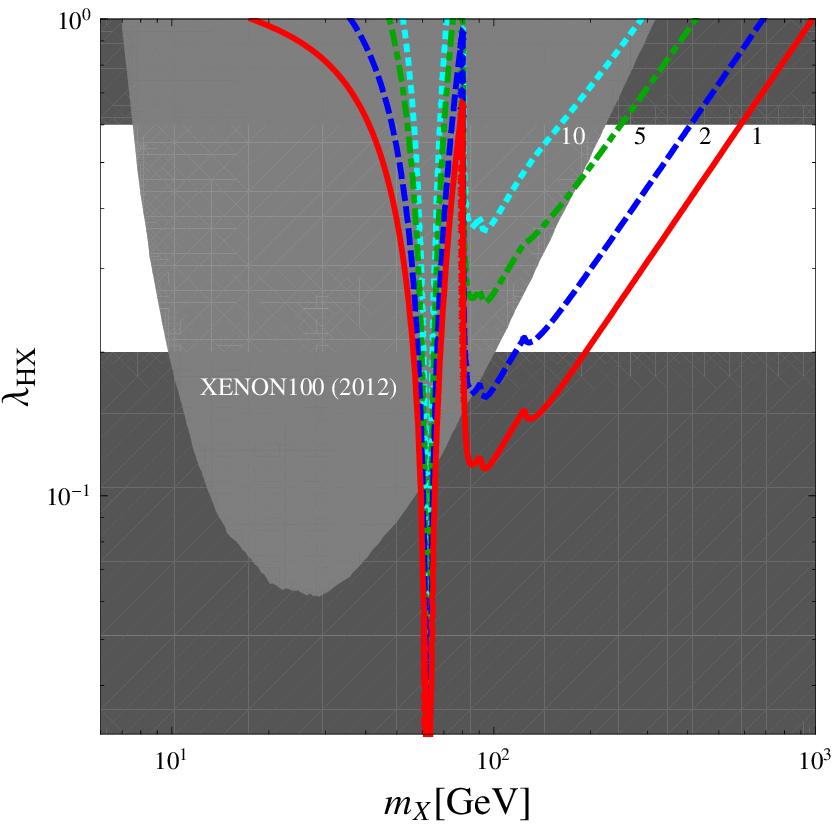}
\caption{
Left: XENON100(2012) bound on the kinetic mixing parameter $\epsilon$ as a function of $m_X$ for $\alpha_X$ 
given by the bound value of \eq{DMstructure-const}.
Right: XENON100 (2012) bound and contours of $\langle \sigma v \rangle_{\rm ann}^X 
/ \langle \sigma v \rangle_{\rm ann}^{\rm th}$ in ($\lambda_{HX}$, $m_X$) plane. 
$\langle \sigma v \rangle_{\rm ann}^{\rm th} \simeq 2 \times 10^{-36} {\rm cm}^2$ is the thermally 
averaged annihilation cross section giving the correct amount of dark matter relic density from thermal freeze-out. The gray region is excluded by the recent result from XENON100 \cite{Aprile:2012nq}. 
The lower and upper dark gray region is excluded by the vacuum stability of Higgs potential and perturbativity of couplings, respectively. 
The colored lines correspond to $\langle \sigma v \rangle_{\rm ann}^X / \langle \sigma v 
\rangle_{\rm ann}^{\rm th} = 1, 2, 5, 10$ from bottom to top.}
\label{fig:lHX-Xenon-bound}
\end{figure}

In the case of the Higgs portal interaction, the scattering cross section is 
\beq \label{sp-SI}
\sigma_{\mathcal  {N}, H}^{\rm SI} = \frac{1}{\pi} m_{\rm r}^2 \l[  \frac{1}{8} \lambda_{HX} \frac{m_{\mathcal N}}{m_X m_H^2} \left( \sum_{q=u,d,s} f_{Tq}^{\mathcal N} + \frac{2}{27} \sum_{q=t,b,c} f_{TG}^{\mathcal N} \right) \r]
\eeq
where $f_{Tq}^{\mathcal N}$ and $f_{TG}^{\mathcal N}$ being hadronic matrix elements with a scalar.
In \eq{sp-SI}, we take $\l( \cdots \r) = 0.326$ \cite{Young:2009zb}.
Currently, the strongest bound on $\sigma_p^{\rm SI}$ comes from XENON100 direct search experiment \cite{Aprile:2012nq} which has $E_{\rm r}^{\rm T} = 6.6 \keV$. 
Fig.~\ref{fig:lHX-Xenon-bound} shows how the kinetic mixing (left panel) and Higgs portal coupling (right panel) are limited by the experiment (gray region).
%
%
Also, depicted are thermally averaged annihilation cross sections (colored lines) and bounds 
from vacuum stability and perturbativity (dark gray regions).
In the left panel of the Fig.~2, we notice that, if small scale anomalies of structure formation are to be explained by the dark matter self-interaction, XENON100 direct search experiment constrains strongly the kinetic mixing as 
\beq
\epsilon \lesssim 10^{-9} - 10^{-4} \quad {\rm for} \quad 6 \GeV \lesssim m_X \lesssim 1 \TeV.
\eeq
From the right panel of Fig.~2, we also notice that direct search experiments already excluded $m_X \lesssim 80 \GeV$ except for the narrow resonance band around $m_X \simeq m_h/2$. 
In addition, for $m_X = \mathcal{O}(10^{2-3}) \GeV$, $\langle \sigma v \rangle_{\rm ann}^X$ is constrained to be within the range of
\beq \label{sv-dd-const}
1 \leq \frac{\langle \sigma v \rangle_{\rm ann}^X}{\langle \sigma v \rangle_{\rm ann}^{\rm th}} \lesssim 5.
\eeq

\subsection{Indirect Signatures}
\label{sec:indirect-dec}
The dark interaction and kinetic mixing in our model should be highly suppressed as described in previous sections.
In addition, since $\alpha_X \lesssim 10^{-4}$ for $m_X \lesssim \mathcal{O}(1) \TeV$  (see \eq{DMstructure-const}), Sommerfeld enhancement factor is of order unity.
Hence it is difficult to expect detectable indirect signatures from the annihilation channels 
via dark interaction or kinetic mixing.

The possible indirect detection signatures comes from Higgs portal interactions,   
\[
X X^\dagger \rightarrow H^* \rightarrow f \bar{f}, V V,  \ \ {\rm or} \ \ X X^\dagger \rightarrow H H , 
\]
where $f$ and $V$ are the SM fermions and the weak gauge bosons, respectively.
These processes can produce a sizable continuum spectrum of photons, since the annihilation cross section can be larger than the value for thermal dark matter.
However, the recent data from Fermi LAT $\gamma$-ray search provides upper-bounds on various annihilation channels \cite{,Huang:2012yf}.
In our model, $W^+ W^-$ channel is dominant, and upper-bounded as

\beq
\langle \sigma v \rangle_{\rm ann}^X \lesssim {\rm Br}(XX^\dag \to W^+ W^-)^{-1} \times 2 \times 7.4 \times 10^{-26} {\rm cm}^3 / {\rm sec}.
\eeq
In the allowed region of parameter space, that is, for $m_X = \mathcal{O}(10^{2-3}) \GeV$, we find ${\rm Br}(XX^\dag \to W^+ W^-) \sim 0.5$, and the constraint on the annihilation cross section is nearly same as \eq{sv-dd-const}.

\section{Dark Radiation}  
\label{sec:dark-rad}
Dark photon can contribute to the radiation density of the present universe.
In our scenario dark matter is decoupled from the SM thermal bath at a temperature $T \sim 1 \GeV$ before QCD-phase transition while still in contact with dark photon.
Hence dark matter and dark photon are decoupled from the SM thermal bath simultaneously.
When it is decoupled, the temperature of dark photon is the same as that of photon.
Therefore, the dark photon contribution to the radiation as the extra relativistic neutrino species is 
\beq
\Delta N_{\rm eff} =  \frac{g_{\gamma'}}{(7/8) g_\nu} 
\l( \frac{T_{\gamma,0}}{T_{\nu,0}} \r)^4 \l( \frac{g_{*S}(T_{\gamma, 0})}{
g_{*S}(T_{\gamma, \rm dec})} \r)^{4/3} \simeq 0.08
\eeq
where $g_i$, $T_{i,0}$ and $T_{i, \rm dec}$ are respectively the degrees of freedom, the temperature at present and decoupling of the species, $i$, and $g_{*S}$ is the total SM degrees of freedom associated with entropy. 
We used  $g_{\gamma'} = g_\nu = 2$, $g_{*S}(T_{\gamma, 0}) = 3.9$ and $g_{*S}(T_{\gamma, \rm dec})=75.75$.
The best fit value of observations is \cite{Hinshaw:2012fq}
\beq \label{Neff-obs}
N_{\rm eff}^{\rm obs} = 3.84 \pm 0.40 \ {\rm at} \ 68 \% \ {\rm CL}
\eeq
with SM expectation $N_{\rm eff}^{\rm SM} = 3.046$.
Therefore, in our model the contribution of dark photon to the radiation density 
at present is consistent with observation within about 2-$\sigma$ error, 
slightly improving the SM prediction in the right direction.

\section{Leptogenesis}

The global lepton number is explicitly broken by Majorana mass terms for the RH neutrinos.  
The lightest RH Majorana neutrino $N_1$ can decay into both the SM fields and the DM fields: 
\[
N_1 \rightarrow l_{Li} H^\dagger , \ \ \ \psi X^\dagger .
\]
With nonzero complex phases in $Y_\nu$ and $\lambda_i$ the decay can generate the $\Delta L$, $\Delta \psi$ and $\Delta X$.
Here we consider only $\Delta L$ since the possible asymmetry in the dark sector particles is removed as $\psi$ decays. 

The eventual outcome of leptogenesis via the decay of heavy RHN can be obtained by solving Boltzmann equations which involve effects of wash-out and transfer of the asymmetries between visible and dark sectors.
However, if the narrow-width approximation is valid, we can get much simpler picture.
The narrow-width approximation is valid if $\Gamma_1^2 / M_1 H_1 \ll 1$
and $2 \to 2$ scattering between visible and dark sectors via heavy neutrino is ineffective, hence asymmetries in both sector evolves independently.
In this circumstance, the washout effect on asymmetry is mainly from the inverse decay.
If the washout effect is weak, i.e., ${\rm Br}_i \Gamma_1 / H_1 \ll 1$,
the final asymmetry is directly related to the asymmetry from the decay of RHN.
Otherwise, there can be large reduction of the asymmetry.

\begin{figure}[h] 
\centering
\includegraphics[width=0.5\textwidth]{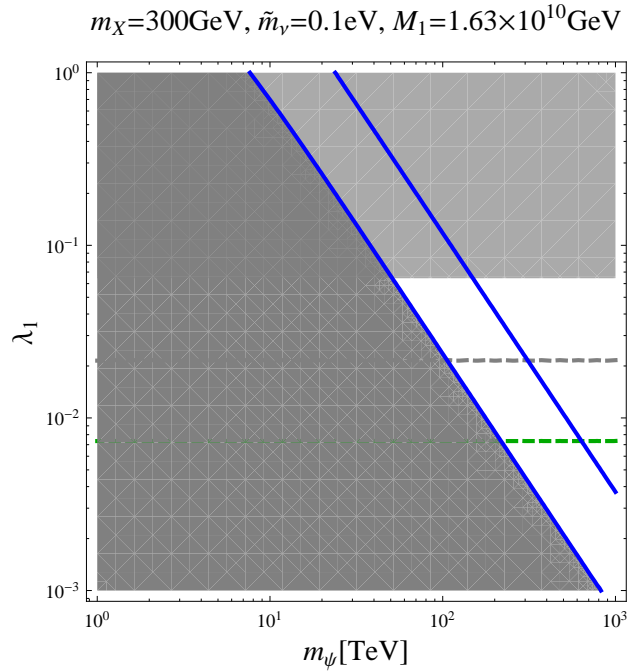}
\caption{Parameter space for right amounts of baryon number asymmetry and dark matter relic density at present. 
We used $m_X = 300 \GeV$, $\tilde{m}_\nu = 0.1 \eV$ and $\sqrt{Y_{\nu 2}^2 M_1 / Y_{\nu 1}^2 M_2} = 1$ corresponding to $M_1 = 1.63 \times 10^{10} \GeV$.
Dark gray region is excluded by XENON100 dark matter direct search experiment. 
In the light gray region, narrow-width approximation is not valid. 
The boarder of the light gray region and the gray dashed line correspond to 
$\lambda_1/ \sqrt{16 \pi \sqrt{M_1/M_\pl}} = 1, 1/3$, respectively.
Below the green line, $Y_{\nu 1} > \lambda_1$ for which our analysis is not valid.
The blue lines correspond to $\langle \sigma v \rangle_{\rm ann}^X / \langle \sigma v \rangle_{\rm ann}^{\rm th} = 1, 5$ from right to left.
}
\label{fig:para-space-narrow-wid-approx}
\end{figure}

For simplicity, we considered the regime of the narrow-width approximation, and found that there is wide range of parameter space where the observed baryon number asymmetry and dark matter relic density can be obtained as shown in Fig.~\ref{fig:para-space-narrow-wid-approx}.
In the figure, dark gray region is excluded by XENON100 direct dark matter search.
Narrow-width approximation is valid in the white region well below the light gray region (e.g., below the dashed gray line).
$\lambda_1 < Y_{\nu 1}$ below the green line.
Although a wider parameter space may be allowed, our analysis of leptogenesis is limited only in the white region bounded by the dashed gray and green lines.
In the region, right amounts of baryon number asymmetry and dark matter relic density can be obtained.

So far, we have considered the lepton number asymmetry in the visible sector that comes from the decay of RH-neutrinos only.
However there is an additional contribution from the late-time decay of $\psi$ which also carries lepton number.
Since the decays of $\psi$ and $\bar{\psi}$ involve a virtual internal line of a Majorana RH-neutrino which decays eventually to a SM lepton and Higgs pair, both of decays produce equal amount of the same-sign lepton number asymmetry in the visible sector.
In addition, there is no dilution of the produced visible sector asymmetry due to inverse decay or transfer to the dark sector, since such processes are kinematically forbidden.
We found that the decay of $\psi$ and $\bar{\psi}$ can be the origin of the baryon number asymmetry in the present universe even though the asymmetry between $\psi$ and $\bar{\psi}$ is absent.

\section{Higgs Inflation}

In order for the leptogenesis described in the previous section to work, 
the temperature of the early universe should be high enough so that the lightest 
RHN can be in thermal equilibrium before it is decoupled.
This condition can be achieved if the reheating temperature of the primordial inflation
is high enough.  An intriguing possibility is so-called Higgs inflation 
\cite{Bezrukov:2007ep,Bezrukov:2010jz} which uses the SM Higgs as the inflaton 
equipped with a large non-minimal gravitational coupling.
As a variant, Higgs-scalar singlet system has been also considered in the literature. 
Modulo the subtle issues of the unitarity problem~\cite{Burgess:2009ea,Lerner:2011it}, 
our model indeed allows inflation along Higgs direction since Higgs potential is stabilized by the 
help of a coupling to the singlet scalar $X$.
The model parameters relevant to inflation are $\lambda_{HX}$, $\lambda_X$ and 
the Higgs quartic coupling in addition to the large non-minimal couplings (say $\xi_i$).
As free parameters, we can adjust $\xi_i$s for given set of quartic couplings while 
satisfying requirements on the inflationary observables under the assumption of the 
positivity of quartic couplings (see \cite{Lebedev:2012zw} for example).
Hence the physics involved in inflation does not pose any new constraint other than ones
described in previous sections if inflation takes place along Higgs direction, and 
the Higgs inflation along with a singlet scalar can be realized.

It turned out that the reheating temperature after Higgs inflation is around 
$\mathcal{O}(10^{13-14}) \GeV$ \cite{Bezrukov:2008ut}.
It is high enough to populate the lightest RHN in thermal bath.
Therefore, Higgs inflation sets the initial condition for the leptogenesis.

\section{Higgs and DM phenomenology at colliders}

The Higgs boson in our model could decay into a pair of scalar DM's through $\lambda_{HX}$ term if kinematically allowed.
However, as shown in Fig.~\ref{fig:lHX-Xenon-bound}, dark matter direct search allows only $m_X \sim m_h /2$ with $\lambda_{HX} \lesssim  10^{-1}$ even though SM Higgs may not suffer from vacuum instability problem.
If it is allowed, the decay rate of Higgs to dark matter is 
\beq
\Gamma_{h \to XX^\dag} = \frac{\lambda_{HX}^2}{128 \pi} \frac{v_H^2}{m_h} \l( 1 - \frac{4 m_X^2}{m_h^2} \r)^{1/2} ,
\eeq
and the signal strength ($\mu$) of SM Higgs searches at collider experiments is given by
\beq
\mu = 1 - \frac{\Gamma_{h \to X X^\dag}}{\Gamma_h^{\rm tot}} 
\eeq
where $\Gamma_h^{\rm tot}$ is the total decay rate of SM Higgs.
Recent results from ATLAS and CMS collaborations are \cite{ATLAS,CMS}
\bea
\mu_{\rm ATLAS} &=& 1.43 \pm 0.21 \quad {\rm for} \ m_h = 125.5 \GeV \ ,
\\
\mu_{\rm CMS} &=& 0.8 \pm 0.14 \quad \hspace{0.5em}  {\rm for} \ m_h = 125.7 \GeV \ .
\eea
Hence the invisible decay of Higgs to dark matter can be consistent with CMS data 
only if  $\lambda_{HX} \ll 0.1$ or $m_X$ is very close to $m_h/2$.
On the other hand, if vacuum stability is imposed, such a small $\lambda_{HX}$ is excluded and only $m_X = \mathcal{O}(10^{2-3}) \GeV$ is allowed.
In this case, the production and decay rate of Higgs boson in our model are exactly 
the same as those of SM Higgs boson. 
It is difficult to discriminate our model from SM in such a case.
In other words, if collider experiments shows any non-SM signature, our model is excluded.
%

\section{Variations of the model}

One can consider a simpler dark sector which contain either $X$ or $\psi$ only in addition to $\hat{B}'_\mu$.  
If $\psi$ is absent and $X, X^\dagger$ are dark matters, the only change relative to our present model is that the current dark matter relic density should come from the thermal freeze-out of $X$-$X^\dag$ annihilation via  $\lambda_{HX}$ interaction.  
Hence the annihilation cross section is fixed to be  the usual value.
If $X$ is absent and $\psi$ is the dark matter, one has to introduce a real SM-singlet 
scalar (say $S$) connecting the dark sector to the SM sector as the model discussed in 
Ref.~\cite{SFDM1}, so that the thermal freeze-out of $\psi$-$\bar{\psi}$ annihilation via the newly introduced interactions of $S$ provides a right amount of 
dark matter at present.  Otherwise $\psi$ and $\bar{\psi}$ would be overproduced 
due to the smallness of $\alpha_X$.  

One could also consider the case the $U(1)_X$ dark symmetry is spontaneously 
broken  by nonzero $\langle \phi \rangle \neq 0$.
In this case there is a singlet scalar from $\phi$ after $U(1)_X$ breaking, which will mix with the SM Higgs boson. 
Therefore there are two Higgs-like neutral scalar bosons after all, and both of
them have signal strengths universally suppressed from the SM value ``1''.  
If $\psi$ is the CDM, one needs a singlet scalar $S$ as a messenger, and this will
mix with the SM Higgs boson (and the remnant from $\phi$).  

In Table~2, we summarize the dark field contents, messengers, the particle identity 
of the dark matter (DM), the amount of dark radiation (DR) and the signal strengths 
of Higgs-like neutral scalar bosons (including the number of them) in various scenarios.
In all cases, there are additional scalar bosons (either $X$ or $\phi$ or both) which 
make Higgs inflation still viable for $m_H = 125$ GeV.   And the Higgs signal strength 
is smaller than ``1'' except for the scalar is the CDM with unbroken $U(1)_X$ dark 
symmetry. Especially $\mu_{i=1,2,(3)} < 1$ for fermion CDM, whether $U(1)_X$ is 
broken or not. Our conclusions on the Higgs signal strength are based on the 
assumption that there is only one Higgs doublet in the model.  If we include 
additional Higgs doublets or triplet Higgs, the Higgs portal would have richer structure,
and the signal strength will change completely and will vary depending on the Higgs 
decay channels. Also it should be possible to have a signal strength for 
$H\rightarrow \gamma\gamma$ channel greater than  ``1''  without difficulty.  

\begin{table}[htdp]
\begin{tabular}{|c|c|c|c|c|c|}
\hline
Dark sector fields & $U(1)_X$ & Messenger & DM & Extra DR & $\mu_i$  
\\   \hline  
$\hat{B}'_\mu , X , \psi$ & Unbroken & $H^\dagger H  , \hat{B}'_{\mu\nu} \hat{B}^{\mu\nu} , N_R$ 
& $X$ & $\sim 0.08$ & $1~(i=1)$
\\
$\hat{B}'_\mu , X$ & Unbroken & $H^\dagger H , \hat{B}'_{\mu\nu} \hat{B}^{\mu\nu}$ 
& $X$ & $\sim 0.08$ & $1 ~(i=1)$
\\
$\hat{B}'_\mu , \psi$ & Unbroken & $H^\dagger H  , \hat{B}'_{\mu\nu} \hat{B}^{\mu\nu} ,  S$ 
& $\psi_X$ & $\sim 0.08$ & $< 1~(i=1,2)$
\\
\hline 
$\hat{B}'_\mu , X , \psi , \phi$ & Broken & 
$H^\dagger H  , \hat{B}'_{\mu\nu} \hat{B}^{\mu\nu} , N_R$ 
& $X$ or $\psi$ & $\sim 0$ & $< 1~(i=1,2)$
\\
$\hat{B}'_\mu , X , \phi$ & Broken 
& $H^\dagger H , \hat{B}'_{\mu\nu} \hat{B}^{\mu\nu}$ & $X$ & $\sim 0$ & $< 1 ~(i=1,2)$
\\
$\hat{B}'_\mu , \psi$ & Broken 
& $H^\dagger H  , \hat{B}'_{\mu\nu} \hat{B}^{\mu\nu} ,  S$ 
& $\psi$ & $\sim 0$ & $~~< 1~(i=1,2,3)$
\\   \hline
\end{tabular}
\caption{Dark fields in the hidden sector, messengers, dark matter (DM), the 
amount of dark radiation (DR), and the signal strength(s) of the $i$ scalar boson(s) 
($\mu_i$)  for unbroken or spontaneously broken (by $\langle \phi \rangle \neq 0$) 
$U(1)_X$ models considered 
in this work.  The number of Higgs-like neutral scalar bosons could be 1,2 or 3, 
depending on the  scenarios. }
\label{default}
\end{table}%

\section{Conclusion}
In this talk, we showed that, if the dark matter stability is guaranteed by an unbroken 
local dark symmetry with nonzero dark charge, renormalizable portal interactions of 
the RH neutrinos and SM Higgs ($H^\dagger H$) fix the minimal field contents of 
dark sector (a scalar $X$ and fermion $\psi$ as well as massless dark photon 
$\hat{B}'_\mu$) and allow very rich physics without conflicting with various 
phenomenological, astrophysical and cosmological constraints coming from the 
existence of massless dark photon.

The model parameters are highly constrained.
Especially, due to small and large scale structure formation the dark fine structure constant of the unbroken local dark symmetry is constrained to be $\alpha_X \lesssim 10^{-5} - 10^{-4}$ for $\mathcal{O}(10^{2-3}) \GeV$ scale mass of dark matter.  
Interestingly, when it is close to its upper-bound, the dark matter self-interaction can be the solution to the core/cusp and ``too big to fail'' problems of small scale structure in collisionless CDM scenarios.  
The smallness of dark interaction could cause a danger of dark matter over-abundance, but this potential danger is removed by RH neutrino portal which allows $\psi$ to decay to $X$, and Higgs portal which allows efficient dilution of the stable dark matter $X$ 
to get a right amount of dark matter relic density.   
All these nice features are consequences of local dark gauge symmetry, and the assumption that all the SM singlet operators being portals to the dark sector.  
The RH neutrino portal also allows production of dark sector asymmetry as leptogenesis in type-I seesaw model does in 
the visible sector, but the dark sector asymmetry eventually disappears as $\psi$ decays and does not play any significant role.
However, one should note that in our scenario eventual relic density of dark matter can be determined by thermal or non-thermal freeze-out, depending on the temperature when $\psi$ decays to $X$.  This allows wider range of dark matter 
annihilation cross section. 
Additionally, depending on $\alpha_X$ and the mass of $\psi$, the decays $\psi$ and $\bar{\psi}$ can be the origin of the present baryon number asymmetry irrespective of the possible asymmetry between $\psi$ and $\bar{\psi}$. 

The Higgs portal interaction to dark scalar $X$ cures instability problem of the SM Higgs potential by loop effect.  So, by introducing large non-minimal gravitational couplings to scalar fields, it becomes possible to realize Higgs inflation whose high enough reheating temperature sets the initial condition for  leptogenesis in our model.
The portal interactions also make the dark sector be accessible by direct and/or indirect searches, which are consistent with the current bounds from various terrestrial experiments and observations in the sky.  

It turned out that the contribution of dark photon to the radiation density at present is about 8\% of the energy density of a massless neutrino. 
It is rather small, and still consistent with the present observation within $2$-$\sigma$ error. 
The smallness is originated from the fact that dark photon couples only to the dark sector fields and dark matter is decoupled from SM particles before QCD-phase transition.
 
Our model is a sort of the minimal model which has inflation, leptogenesis, absolutely stable dark matter and dark radiation, as well as seesaw mechanism for neutrino masses and mixings, modulo some variations described at section~8.
It can be a good alternative to the SM, covering the phenomenological short-comings of the SM.
The basic principles for the model building were the local gauge symmetry working on the dark sector too, and the assumption of the SM singlet operators being portals to the dark sector. From these simple principles,  we could derive very rich physics results that are fully consistent with all the observations made so far.  
It is interesting to note that the Higgs property measurements will strongly constrain our model, since we predict that  the Higgs signal strength should be equal to or less than ``1" for all the decay channels of Higgs  boson. 

\bibliographystyle{plain}

\begin{thebibliography}{9}

\bibitem{leptogenesis}
M.~Fukugita and T.~Yanagida,
  Phys.\ Lett.\ B {\bf 174}, 45 (1986).

\bibitem{R2}
  A.~A.~Starobinsky,
  Phys.\ Lett.\ B {\bf 91}, 99 (1980).
  
\bibitem{Bezrukov:2007ep} 
  F.~L.~Bezrukov and M.~Shaposhnikov,
  Phys.\ Lett.\ B {\bf 659}, 703 (2008)
  [arXiv:0710.3755 [hep-th]].


\bibitem{Ackermann:2012qk} 
  M.~Ackermann {\it et al.}  [LAT Collaboration],
  Phys.\ Rev.\ D {\bf 86}, 022002 (2012)
  [arXiv:1205.2739 [astro-ph.HE]].
  

\bibitem{Baek:2013qwa} 
  S.~Baek, P.~Ko and W.~-I.~Park,
  JHEP {\bf 1307}, 013 (2013)
  [arXiv:1303.4280 [hep-ph]].


\bibitem{Hur:2011sv} 
  T.~Hur and P.~Ko,
  Phys.\ Rev.\ Lett.\  {\bf 106}, 141802 (2011)
  [arXiv:1103.2571 [hep-ph]];



\bibitem{Baek:2011aa} 
  S.~Baek, P.~Ko and W.~-I.~Park,
  JHEP {\bf 1202}, 047 (2012)
  [arXiv:1112.1847 [hep-ph]].

\bibitem{Baek:2012se} 
  S.~Baek, P.~Ko, W.~-I.~Park and E.~Senaha,
  JHEP {\bf 1305}, 036 (2013)
  [arXiv:1212.2131 [hep-ph]].


\bibitem{Ackerman:2008gi} 
  L.~Ackerman, M.~R.~Buckley, S.~M.~Carroll and M.~Kamionkowski,
  Phys.\ Rev.\ D {\bf 79}, 023519 (2009)
  [arXiv:0810.5126 [hep-ph]].

\bibitem{MiraldaEscude:2000qt} 
  J.~Miralda-Escude,
  astro-ph/0002050.

\bibitem{Randall:2007ph} 
  S.~W.~Randall, M.~Markevitch, D.~Clowe, A.~H.~Gonzalez and M.~Bradac,
  Astrophys.\ J.\  {\bf 679}, 1173 (2008)
  [arXiv:0704.0261 [astro-ph]].




\bibitem{Vogelsberger:2012ku} 
  M.~Vogelsberger, J.~Zavala and A.~Loeb,
  arXiv:1201.5892 [astro-ph.CO].


\bibitem{Oh:2010ea} 
  S.~-H.~Oh, W.~J.~G.~de Blok, E.~Brinks, F.~Walter and R.~C.~Kennicutt, Jr,
  arXiv:1011.0899 [astro-ph.CO].


%


\bibitem{BoylanKolchin:2011dk} 
  M.~Boylan-Kolchin, J.~S.~Bullock and M.~Kaplinghat,
  Mon.\ Not.\ Roy.\ Astron.\ Soc.\  {\bf 422}, 1203 (2012)
  [arXiv:1111.2048 [astro-ph.CO]].

\bibitem{Feng:2009mn}
  J.~L.~Feng, M.~Kaplinghat, H.~Tu and H.~-B.~Yu,
  JCAP {\bf 0907} (2009) 004
  [arXiv:0905.3039 [hep-ph]].

\bibitem{Hofmann:2001bi} 
  S.~Hofmann, D.~J.~Schwarz and H.~Stoecker,
  Phys.\ Rev.\ D {\bf 64}, 083507 (2001)
  [astro-ph/0104173].

\bibitem{Alekhin:2012py} 
  S.~Alekhin, A.~Djouadi and S.~Moch,
  Phys.\ Lett.\ B {\bf 716}, 214 (2012)
  [arXiv:1207.0980 [hep-ph]].



  
\bibitem{Lebedev:2012zw} 
  O.~Lebedev,
  Eur.\ Phys.\ J.\ C {\bf 72}, 2058 (2012)
  [arXiv:1203.0156 [hep-ph]].
  
\bibitem{EliasMiro:2012ay} 
  J.~Elias-Miro, J.~R.~Espinosa, G.~F.~Giudice, H.~M.~Lee and A.~Strumia,
  JHEP {\bf 1206}, 031 (2012)
  [arXiv:1203.0237 [hep-ph]].

\bibitem{Baek:2012uj} 
  S.~Baek, P.~Ko, W.~-I.~Park and E.~Senaha,
  arXiv:1209.4163 [hep-ph].


\bibitem{Rodejohann:2012px} 
  W.~Rodejohann and H.~Zhang,
  JHEP {\bf 1206}, 022 (2012)
  [arXiv:1203.3825 [hep-ph]].


\bibitem{Lewin:1995rx} 
  J.~D.~Lewin and P.~F.~Smith,
  Astropart.\ Phys.\  {\bf 6}, 87 (1996).

\bibitem{Young:2009zb} 
  R.~D.~Young and A.~W.~Thomas,
  Phys.\ Rev.\ D {\bf 81}, 014503 (2010)
  [arXiv:0901.3310 [hep-lat]].

\bibitem{Aprile:2012nq} 
  E.~Aprile {\it et al.}  [XENON100 Collaboration],
  Phys.\ Rev.\ Lett.\  {\bf 109}, 181301 (2012)
  [arXiv:1207.5988 [astro-ph.CO]].
  
\bibitem{Huang:2012yf} 
  X.~-Y.~Huang, Q.~Yuan, P.~-F.~Yin, X.~-J.~Bi and X.~-L.~Chen,
  arXiv:1208.0267 [astro-ph.HE].

\bibitem{Hinshaw:2012fq} 
  G.~Hinshaw, D.~Larson, E.~Komatsu, D.~N.~Spergel, C.~L.~Bennett, J.~Dunkley, M.~R.~Nolta and M.~Halpern {\it et al.},
  arXiv:1212.5226 [astro-ph.CO].

\bibitem{Bezrukov:2010jz} 
  F.~Bezrukov, A.~Magnin, M.~Shaposhnikov and S.~Sibiryakov,
  JHEP {\bf 1101}, 016 (2011)
  [arXiv:1008.5157 [hep-ph]].
  
%
%
%
%

\bibitem{Burgess:2009ea} 
  C.~P.~Burgess, H.~M.~Lee and M.~Trott,
  JHEP {\bf 0909}, 103 (2009)
  [arXiv:0902.4465 [hep-ph]].

%
%


\bibitem{Lerner:2011it} 
  R.~N.~Lerner and J.~McDonald,
  JCAP {\bf 1211}, 019 (2012)
  [arXiv:1112.0954 [hep-ph]].
    
\bibitem{Bezrukov:2008ut} 
  F.~Bezrukov, D.~Gorbunov and M.~Shaposhnikov,
  JCAP {\bf 0906}, 029 (2009)
  [arXiv:0812.3622 [hep-ph]].


\bibitem{ATLAS}
  https://atlas.web.cern.ch/Atlas/GROUPS/PHYSICS/CONFNOTES/ATLAS-CONF-2013-014/
\bibitem{CMS}
  http://cms.web.cern.ch/org/cms-papers-and-results  

\bibitem{SFDM1}
  S.~Baek, P.~Ko and W.~-I.~Park,
  JHEP {\bf 1202} (2012) 047
  [arXiv:1112.1847 [hep-ph]].




\end{thebibliography}

\end{document}